\documentclass[useAMS,usenatbib]{mn2e}
\usepackage{times}
\usepackage{graphicx}
\usepackage{amsmath}
\usepackage{natbib}
\bibpunct{(}{)}{;}{a}{}{,}
                                                    
\title[VLT photometry globular cluster systems in Antlia]{VLT 
photometry in the Antlia Cluster: the giant ellipticals NGC\,3258 and NGC\,3268 
and their globular cluster systems\thanks{Based on observations carried out at the
European Southern Observatory, Paranal (Chile). Program 71.B-0122(A).}}
\author[L. P. Bassino,  T. Richtler and B. Dirsch]{Lilia P. Bassino$^{1}$
\thanks{E-mails:\,lbassino@fcaglp.unlp.edu.ar\,(LPB); tom@mobydick.cfm.udec.cl\,(TR); 
borischacabuco@yahoo.co.uk\,(BD)}, Tom Richtler$^{2\star}$ and Boris Dirsch$^{2\star}$\\
$^{1}$Facultad de Ciencias Astron\'omicas y Geof\'{\i}sicas,
       Universidad Nacional de La Plata,
       Paseo del Bosque S/N, 1900-La Plata,\\
       Argentina; and IALP-CONICET, \\
$^{2}$Universidad de Concepci\'on, Departamento de F\'{\i}sica,
      Casilla 160-C, Concepci\'on, Chile (contact address for B.D.)}

\begin{document}

\date{Accepted . Received ; in original form }

\pagerange{\pageref{firstpage}--\pageref{lastpage}} \pubyear{}

\maketitle

\label{firstpage}

\begin{abstract}
We present a deep VLT photometry in the regions surrounding the two dominant  
galaxies of the Antlia cluster, the giant ellipticals NGC\,3258 and  
NGC\,3268. 
We construct the luminosity functions of their globular 
cluster systems (GCSs) and determine their distances through 
the turn-over magnitudes. These distances are in good agreement with those obtained 
by the SBF method. There is some, but not conclusive, evidence that the distance 
to NGC\,3268 is larger by several Mpc. The GCSs colour distributions are bimodal 
but the brightest globular clusters (GCs) show a unimodal distribution 
with an intermediate colour peak.  
The radial distributions of both GCSs are well fitted by de\,Vaucouleurs 
laws up to 5\,arcmin. 
Red GCs present a steeper radial density profile than the blue GCs,  
and follow closely the galaxies' brightness profiles. Total GC populations are 
estimated to be about $6000\pm150$\,GCs in NGC\,3258 and 
$4750\pm150$\,GCs in NGC\,3268. 
We discuss the possible existence of GCs in a field located between the 
two giant galaxies (intracluster GCs). Their luminosity functions and number densities 
are consistent with the two GCSs overlapping in projection.      
\end{abstract}

\begin{keywords}
galaxies: individual: NGC 3258, NGC 3268
-- galaxies: clusters: general -- galaxies: elliptical and lenticular, cD
-- galaxies: star clusters -- galaxies: photometry -- galaxies: haloes
\end{keywords}

%=====================================================================
\section{Introduction}

The Antlia galaxy cluster has long been overlooked in optical studies despite the fact that
it is the nearest cluster after Virgo and Fornax, with a comparable number of members
and total mass as the latter \citep{fer90,ped97,nak00}. Its central part consists of two subgroups, 
each dominated by one of the giant elliptical galaxies NGC\,3258 and NGC\,3268. 
This particular structure makes it an even more interesting target, as evidence 
for interactions between the galaxies in the central cluster region  
may emerge.
However, large differences in the radial velocities between NGC\,3268 and several 
close and bright neighbours suggest a considerable structural depth.
We have performed the first CCD study of the stellar population in NGC\,3258 and 
NGC\,3268 \citep*[][hereafter Paper\,{\sc i}]{dir03a}, where the existing literature
on the Antlia cluster is summarized.

Afterwards, our study of the galaxy content of this cluster \citep{smi07}, revealed the 
presence of numerous low surface brightness galaxies, that had not been identified in a 
former photographic search carried out by \citet{fer90}. 

In Paper\,{\sc i}, the luminosity and colour profiles of NGC\,3258 and NGC\,3268, 
and of their globular cluster systems (GCSs), were 
studied on the basis of wide-field Washington photometry. Both GCSs show   
bimodal colour distributions, but small number statistics prevented the detection  
of any difference between the radial profiles of the two globular cluster (GC) 
subpopulations. 
Unfortunately, these data 
were not deep enough to reach the turn-over magnitudes of the respective 
GCS luminosity functions, which can be used as distance indicators, and  
it was not possible to estimate new distances. The SBF distances determined 
by \citet{ton01} are the only ones available, besides a distance estimation via the 
Hubble flow \citep{hop85}. 

A recent study of GCSs in eight brightest cluster galaxies by  
\citet{har06} includes both Antlia ellipticals. On the basis 
of ($B,I$) photometry obtained with the ACS/WFC camera from the Hubble Space 
Telescope (field of view of  $\approx 3.4 \times 3.4$ arcmin$^2$), 
they focus on the two-colour data and metallicity distributions. 
Harris et al. confirm our results from Paper\,{\sc i} that the GC colour 
distributions in these Antlia galaxies are bimodal. They found a trend 
in the sense that brightest blue GCs seem to become redder with 
increasing luminosity. With regard to the radial projected distribution, 
Harris et al. also show that red GCs are more centrally concentrated than blue ones. 

In this new investigation with VLT data, we determine GC luminosity functions (GCLFs) and, 
through their turn-overs, the distance estimates for the dominant Antlia ellipticals. 
We study the characteristics of the different GC subpopulations 
in each galaxy, and compare them to the galaxies' light profiles. 
The total GC populations are also calculated. In addition, 
the GCSs near the giant elliptical galaxies 
are compared to the cluster population in a field further away, 
about 100\,kpc from both galaxies (the `intracluster' field). One goal of this 
comparison is to search for intracluster GC candidates, i.e., GCs that may be  
unbound to a parent galaxy, but are instead moving freely in the potential 
well of the cluster \citep[e.g.][]{wes95}. 

This paper is organized as follows: Section 2 describes the observations,  
the adopted criteria for the GC candidates' selection, and 
the completeness and reddening corrections. In Section 3 we analyse the results 
from the four observed fields and perform the distance calculations. 
Section 4 deals with the galaxies' properties, and Sections 5 and 6 with 
those of their GCSs.
A discussion of the results is presented in Section 7 and, lastly, the conclusions 
and a summary are provided in Section 8.

%=====================================================================
\section{Observations and reductions}

\subsection{Observations}

Bessel $V$ and $I$ imaging was obtained for four fields in the Antlia cluster 
during 2003 March 27--28, with FORS1 at the VLT UT1 (Antu) telescope (Cerro Paranal, 
Chile). This camera is equipped with a $2048 \times 2048$  pixels$^2$ CCD chip, 
which provides an image scale of 0.2\,arcsec\,pixel$^{-1}$ and a field of view 
of $6.8 \times 6.8$ arcmin$^2$ (about $60 \times 60$ kpc$^2$ at the Antlia 
distance). 

\begin{figure}
\includegraphics[width=84mm]{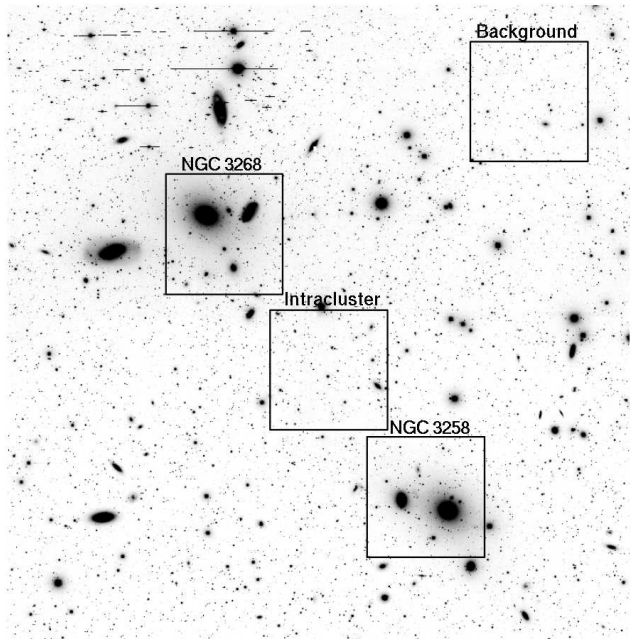}
 \caption{The positions of the four observed FORS1 fields are shown. The underlying image
is the $R$ MOSAIC image that has been used in Paper\,{\sc I}. North is up and east to the left.}
 \label{fig:fields} 
\end{figure}

The positions of the fields are shown in Fig.\,\ref{fig:fields} and basic data are listed in 
Table\,\ref{tab:fields}.  The labels of the fields (see Fig.\,\ref{fig:fields}) were 
selected as follows: `NGC\,3258' and `NGC\,3268' for those that are located on the dominant 
galaxies, `intracluster' for the field placed in between them, and `background' for the field 
located close to the border of the MOSAIC field, used to correct for the contamination by 
the background. For the four fields, three images with exposure times of 100\,/\,200\,s 
each plus five images with exposure times of 300\,/\,700\,s each 
were obtained in the $V$\,/\,$I$ bands, respectively. In all cases, short exposures of 10\,s 
were also taken to avoid saturation problems. The seeing was excellent (Table\,\ref{tab:fields}).

\begin{table*}
\begin{tabular}{cccccc}
\hline
Field & RA(J2000) & DEC(J2000) & Date & Seeing & $E(B-V)$\\
\hline 
NGC\,3258 	& $10^h29^m00.0^s$ & $-35\degr35\arcmin28\farcs3$ & 3/27/2003 & $0\farcs53$ & 0.084 \\ 
NGC\,3268 	& $10^h29^m54.5^s$ & $-35\degr20\arcmin27\farcs4$ & 3/27/2003 & $0\farcs54$ & 0.101 \\
Intracluster 	& $10^h29^m27.6^s$ & $-35\degr28\arcmin20\farcs6$ & 3/28/2003 & $0\farcs64$ & 0.090 \\
Background 	& $10^h28^m31.7^s$ & $-35\degr12\arcmin51\farcs1$ & 3/28/2003 & $0\farcs56$ & 0.091 \\
\hline
\end{tabular}
\caption{Basic data of the observations. Position of the field: 
cols.\,(2) and (3), date of observation: col.\,(4), seeing on the final, 
combined $V$ image: col.\,(5), and reddening towards the center of the field 
according to \citet*{sch98}: col.\,(6).} 
         
\label{tab:fields}
\end{table*}

\subsection{Photometry and point sources selection}

The photometry has been done with DaoPhot\,{\sc ii} within IRAF, 
with the tasks DAOFIND, PSF and ALLSTAR. In the final $V$ and $I$ images, 
a second order variable PSF was derived using an average of 30 evenly distributed 
stars per frame. The aperture corrections were estimated for each field and each band. 
The point sources selection was performed using the $\chi$ and sharpness parameters 
calculated by ALLSTAR. 

\begin{figure}
\includegraphics[width=84mm]{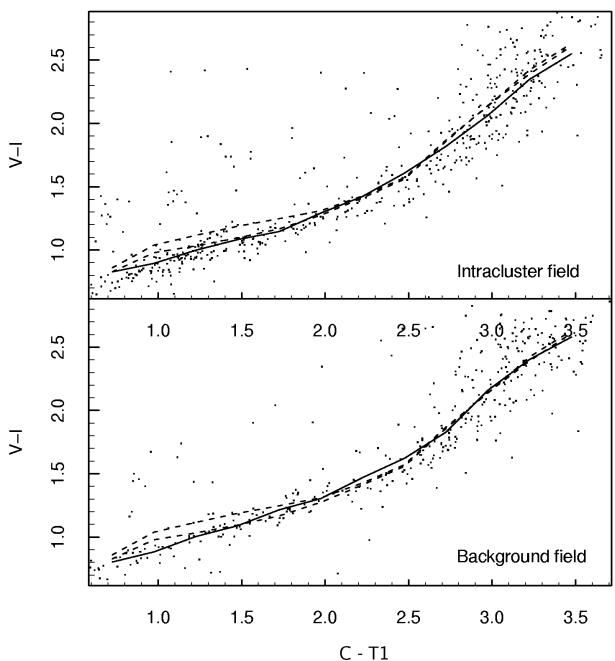}
 \caption{Two-colour diagrams for the `intracluster' and `background' fields, for the
	point sources for which $V,I$ photometry from this paper and $C,T1$ photometry 
	from Paper\,{\sc i} are available. 
        The solid lines show the median relation in the respective fields. The dashed lines 
        show the relation in the NGC\,3258 and NGC\,3268 fields, observed 
        under photometric conditions.
	The comparison for $(C-T1)>2$ verifies that in both fields the 
        applied calibration is correct, while for $(C-T1)<2$ it is 
        affected by the different properties of the GCSs in the four fields (see text).}  
 \label{fig:checkcalib} 
\end{figure}

The first night, in which NGC\,3258 and NGC\,3268 were observed,
was photometric and the data was calibrated 
with the zero-points, airmasses and colour coefficients provided by 
the European Southern Observatory (ESO). As standard stars used by ESO are 
taken from \citet{lan92}, we obtain magnitudes and colours in the 
Johnson/Cousins system.
For the second night, no zero-point is given and the 
flux measurements (available on the ESO-Web site) indicated  
possible presence of clouds until 2 UT. Our
observations started at 3:25 (`intracluster' field) and 5:32 (`background' field),
respectively. Hence, it is highly probable that the night has been photometric
during this time, so we applied the same zero-points as for the first night (the daily
scatter of the zero point is relatively small). The quality of this latter `calibration' 
can be quantified using the Washington ($C,T1$) photometry from Paper\,{\sc i}: the 
two colour diagrams ($V-I$) vs. ($C-T1$) are shown in Fig.\,\ref{fig:checkcalib} for the 
point sources in the `intracluster' and the `background' fields. The median values, indicated
by solid lines, are compared with those from the NGC\,3258 and NGC\,3268 fields shown 
with the dashed lines. 
In the colour range $(C-T1)>2$ the point sources are 
dominated by foreground stars that have identical characteristics in all four fields.
No shift that would indicate a photometric-zero point difference can be seen in this
colour range. The differences for $(C-T1)<2$ are due to the presence/absence of 
GCs and their different properties in the different fields. Similarly, the photometric 
calibration of the $V$ filter has been checked with a ($V-T1$) vs. ($C-T1$) diagram (not shown), 
resulting in the same conclusions.

\begin{figure}
\includegraphics[width=84mm]{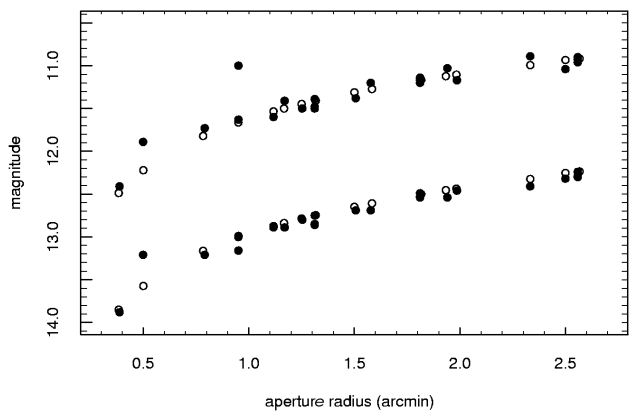}
 \caption{Aperture photometry of NGC\,3268 in $I$ (upper branch) and in $V$ 
	(lower branch) for our data (open circles) and for data from
	the compilation of \citet{pru96} (solid circles).}
 \label{fig:photcheck} 
\end{figure}

The absolute calibration of the first night has been checked using the galaxy aperture 
photometry for NGC\,3258 and NGC\,3268 compiled by \citet{pru96}. 
It is shown for NGC\,3268 in Fig.\,\ref{fig:photcheck}. The difference of the means are 
$\Delta{V} = -0.03\pm0.02$, $\Delta I =0.04\pm0.03$ and $\Delta(V-I)=-0.06\pm0.03$ 
(negative values mean that our measurements are brighter/bluer). For NGC\,3258 we 
find: $\Delta{V} = -0.06\pm0.02$, $\Delta I =0.00\pm0.01$, $\Delta(V-I)=-0.06\pm0.02$. 
The overall agreement is good, however, we measure the galaxies about 0.05 mag bluer 
in ($V-I$). 

We modeled the galaxy light of the two ellipticals with the ELLIPSE task within 
the IRAF/STSDAS/ISOPHOTE package \citep{jed87}. Since the fields are rather crowded, 
we first masked the brightest stars by hand and then used five iterations with a 
3\,$\sigma$ clipping. This procedure can lead to an underestimation of
the total surface brightness, however, the main goal was to obtain a good fitting
galaxy model, that characterizes the radial profile, the ellipticity and the position angle 
(see Section 4). 

\subsection{Completeness}

We performed a standard completeness test by adding $10\times1000$ artificial stars, 
based on the PSF and uniformly distributed, to the four images. We then determined 
the probability that these 
stars were retrieved in an analysis analogous to the one performed on the pure science 
images. We found that the completeness does not vary within the colour range relevant 
for the GCs ($0.75 < V-I <1.4$). It is however, spatially dependent: the completeness 
limit is lower for areas nearer to bright galaxies. This effect is considered later-on.

\begin{figure}
\includegraphics[width=84mm]{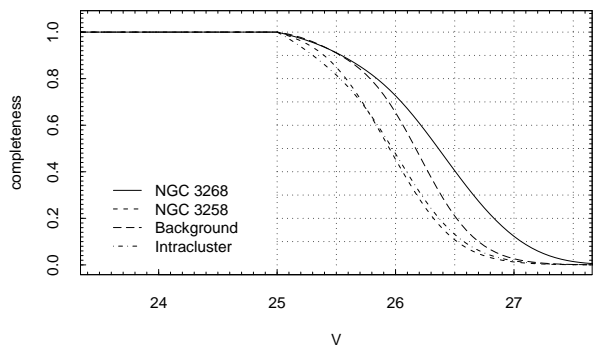}
 \caption{Overall completeness for the four fields.}
 \label{fig:compl} 
\end{figure}

\begin{figure}
\includegraphics[width=84mm]{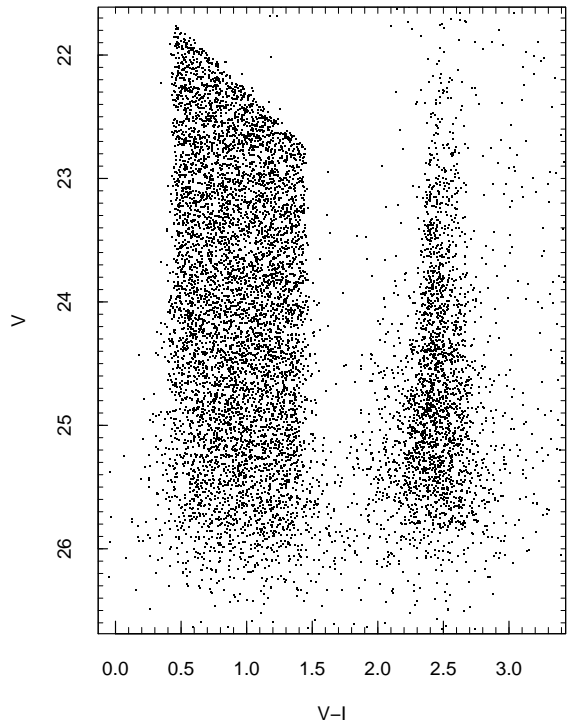}
 \caption{Colour-magnitude diagram of the GCs around NGC\,3258 compared to the 
        CMD of the added
	stars used for the completeness calculation. The former ones are shifted
	in $V-I$ by 1.5 magnitudes redwards to allow an easy comparison.}
 \label{fig:complexample} 
\end{figure}

When one compares the completeness curves shown in Fig.\,\ref{fig:compl} with 
the colour-magnitude diagrams in the next section, a discrepancy becomes apparent, 
particularly in the NGC\,3258 and `intracluster' fields. The completeness appears 
to be dropping faster in these observed fields compared to the prediction of the 
completeness calculations. 
For further illustration, we refer to
Fig.\,\ref{fig:complexample} in which the result of the completeness calculation in
that field is compared to the true (shifted in $V-I$ = 1.5) 
colour-magnitude diagram. 
It appears that the fraction of faint artificial stars is higher than that of the 
true ones. We have no final explanation for this difference. However, the 
artificial stars are not concentrated towards the galaxies, so they will on the average 
have a lower underlying surface brightness and thus will be more sensitive 
to incompleteness effect than the real GCs. 
We want to emphasize that the shape of the completeness curve for low 
luminosities is mainly determined by the used $\chi$ and sha point source selection 
criteria. These criteria 
have been adjusted using the completeness calculations, but apparently true stars in 
the NGC\,3258 field behave {\it worse} when the PSF is fit than the artificial stars. 
We want to note that a similar apparent difference can also be seen in other works,
e.g. \citet*{ost98}.

Due to this uncertainty, we will only use objects brighter than the limit set by an 
70 per cent completeness for the analysis of the GCLFs. Despite this cautious approach, 
we are reaching at least 1.5 mag deeper than in Paper\,{\sc i}.

\subsection{Reddening correction}

We use a conversion factor of $E(V-I)/E(B-V) = 1.2$ 
\citep*{sta96,dea78} and the adopted values for $E(B-V)$ are listed in 
Table\,\ref{tab:fields}. We want to emphasize that the later conversion factor is 
adequate for a Kron-Cousins $I$ filter. 
Later-on we also compare our results to those obtained with Washington 
photometry (Paper\,{\sc i}); for this system the conversion factor is 
$E(C-T1) / E(B-V) = 1.97$ \citep{har77}.

It can be seen from Table\,\ref{tab:fields} that the reddening towards the 
different fields varies from $E(B-V)= 0.08$ to 0.10, which corresponds 
to a range in $E(V-I)= 0.10 - 0.12$. As already stated in Paper\,{\sc i}, 
the IRAS map towards the Antlia cluster is very patchy. In the present work, we are 
using smaller fields than in  Paper\,{\sc i} so we will apply reddening corrections 
according to the individual values listed in Table\,\ref{tab:fields}. However, these 
reddening uncertainties should be kept in mind when comparing results between the 
different fields.
 
%==========================================================
\section{The four fields compared}

\subsection{Colour-magnitude diagrams and colour distributions}

\begin{figure}
\includegraphics[width=84mm]{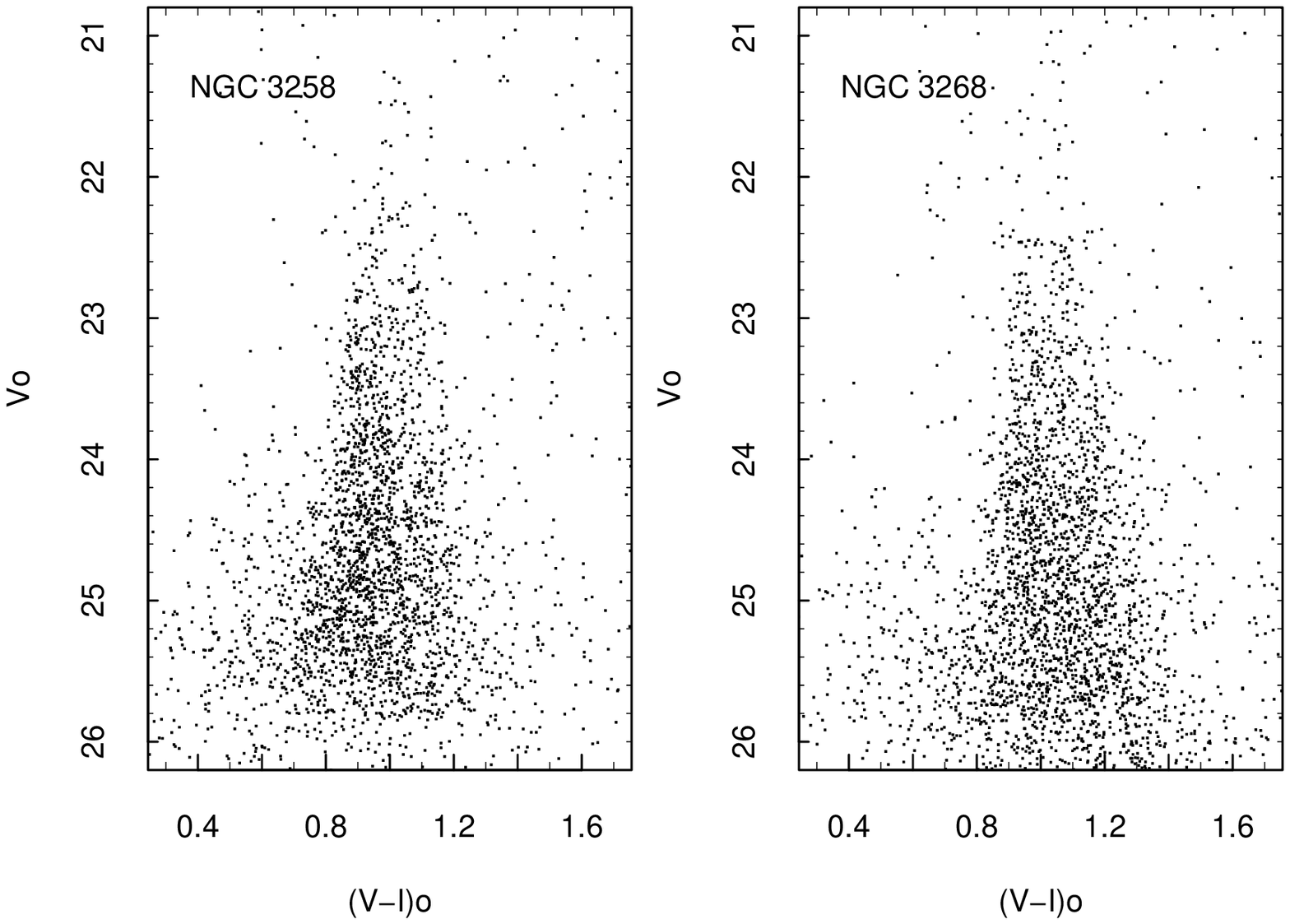}
\includegraphics[width=84mm]{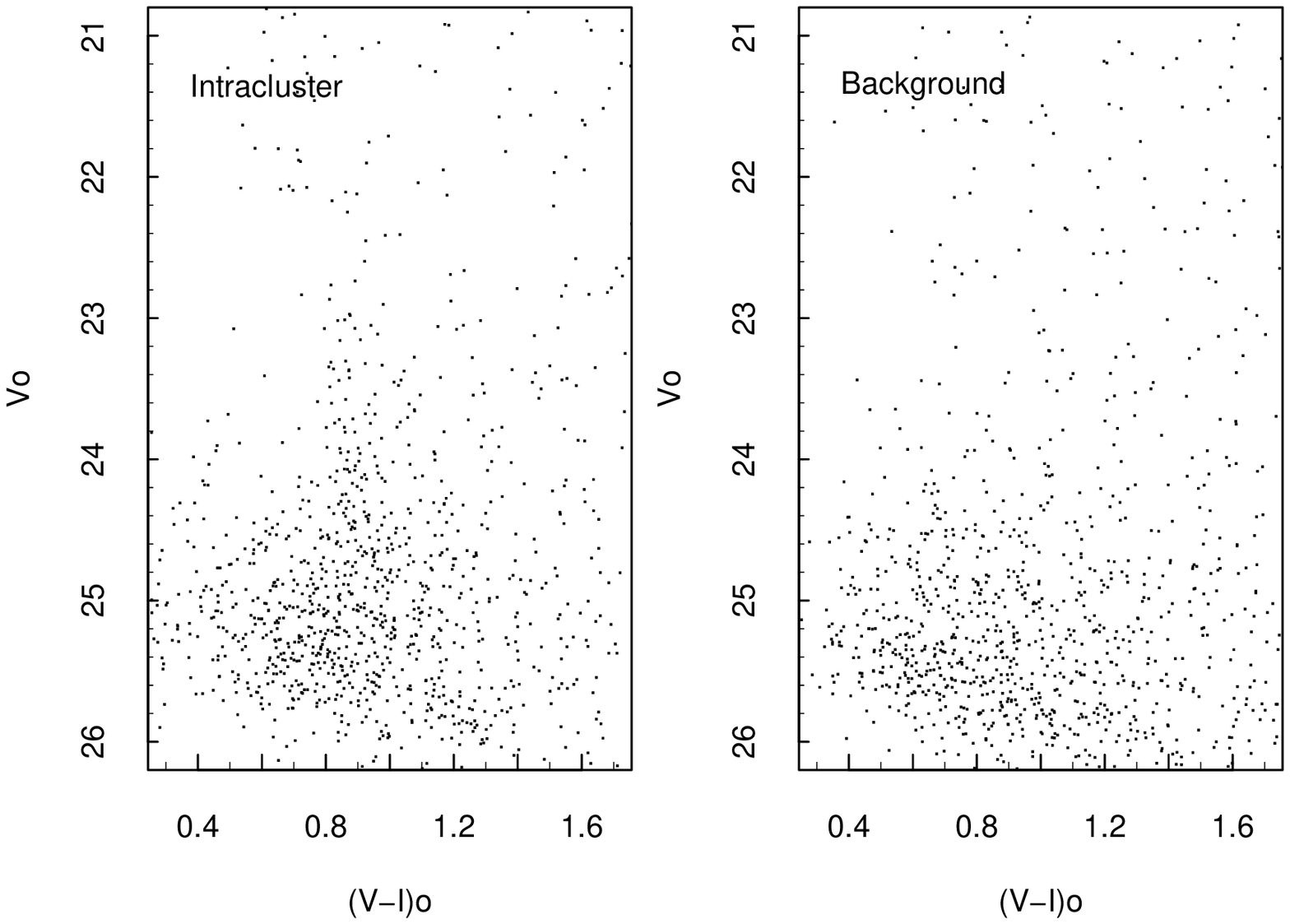}
 \caption{Colour-magnitude diagrams of all point sources in the four studied fields.
The GCs can be seen in the colour ranges $0.75<(V-I)<1.4$ in the NGC\,3258 
and NGC\,3268 fields and $0.8<(V-I)<1.2$ in the `intracluster' field . 
The faint, blue objects that dominate 
the `background' field are predominately background galaxies. 
}
 \label{fig:cmd1} 
\end{figure}

The colour-magnitude diagrams (CMDs), corrected by reddening, of the point sources 
in the four fields 
are shown in Fig.\,\ref{fig:cmd1}.  
The GCs are found in the colour range $0.75 <(V-I)<1.4$. It is 
already discernible that NGC\,3268 has a larger fraction of red clusters than 
NGC\,3258, that extend to redder 
colours. In the `intracluster' field mainly blue GCs are detected.

\begin{figure*}
\includegraphics[width=84mm]{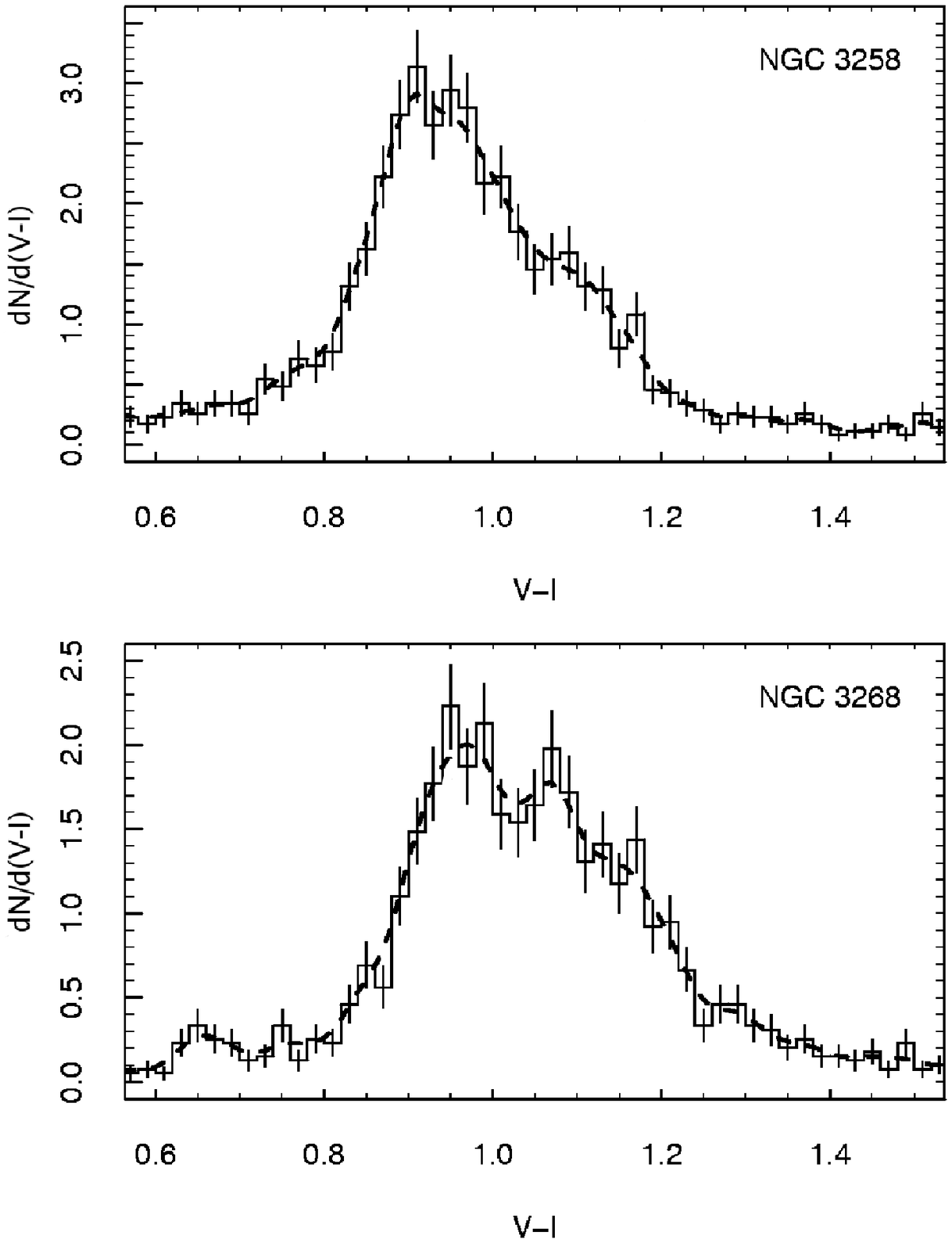}
\includegraphics[width=84mm]{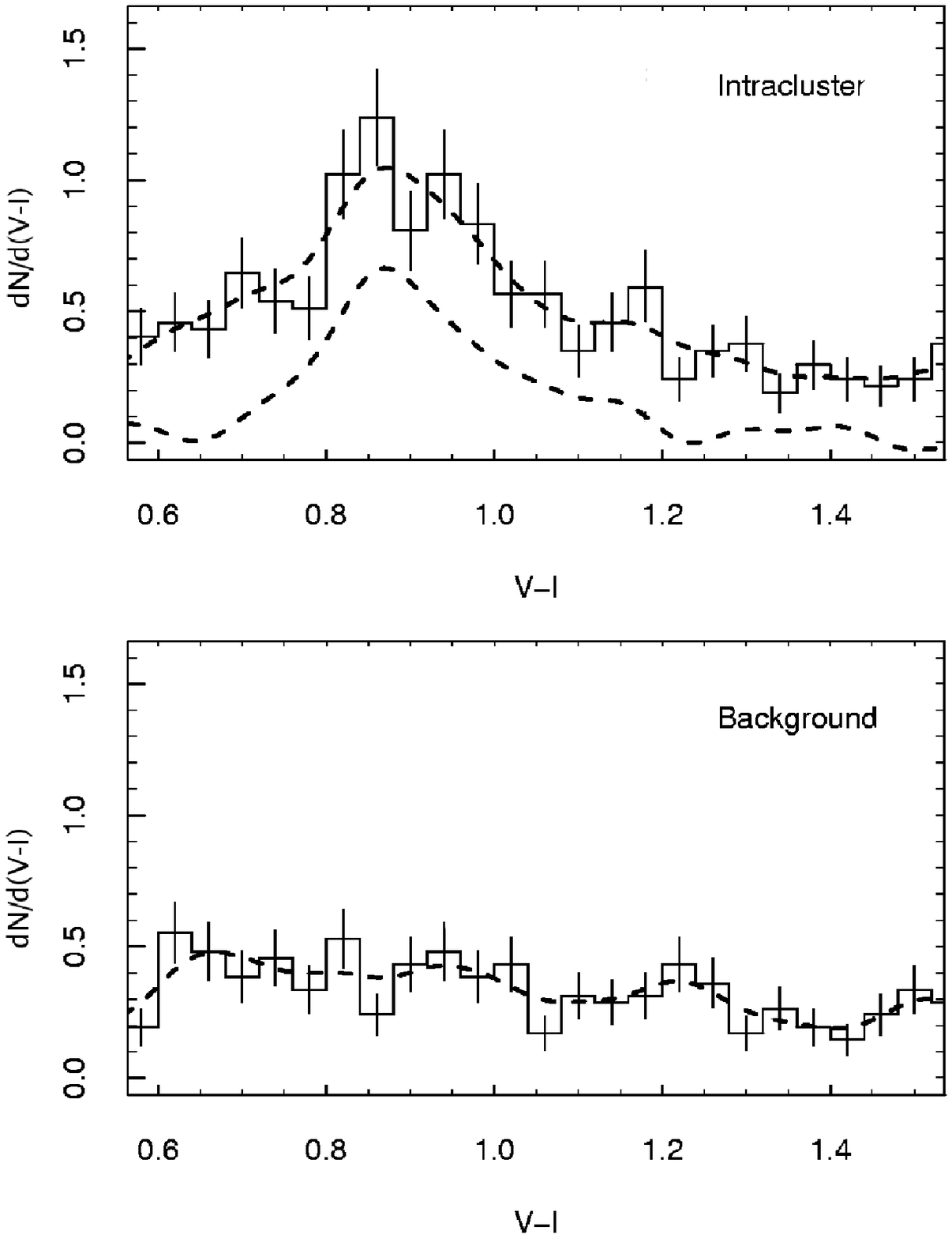}
 \caption{Colour distributions of the points sources brighter than $V$=25.7  
	   for all the studied fields. 
The lower dashed line in the `intracluster' field is the colour distribution 
after the background subtraction.} 
 \label{fig:colour1} 
\end{figure*}

The (reddening corrected) colour distributions of the points sources brighter than
$V$ = 25.7 are shown in Fig.\,\ref{fig:colour1} for the four
fields, with different vertical scales. The mean colour of the GCs in NGC\,3258 
is bluer than of those in NGC\,3268, which can also be seen in Fig.\,\ref{fig:cmd1}  
and in ($C-T1$) (Paper\,{\sc i}). 
In Paper\,{\sc i} a bimodal colour distribution was clearly discernible for the GCSs 
of both elliptical galaxies. It is not as apparent in these new observations, due to 
the roughly two times lower metallicity sensitivity of the ($V-I$) colour 
compared to the ($C-T1$) colour, though bimodality is already visible in Fig.\,\ref{fig:cmd1}.  
A KMM test based on the code of \citet*{ash94} gives $p$-values smaller 
than 0.001 for the colour distributions of both GCSs. These low $p$-values indicate  
that two Gaussians give a better fit the colour distributions than a single Gaussian. 
The bimodality in both GCS colour distributions is also clearly established by the 
($B-I$) colour histograms depicted by \citet{har06} in their fig. 9.
The results of a two-Gaussian fit, based on a nonlinear least-squares code, 
are tabulated in Table\,\ref{tab:kmm}. 
We recall that, in massive early-type 
galaxies, the most common values are $(V-I) = 0.95 \pm0.02$ and  
$1.18 \pm0.04$ for the blue and red peaks, respectively (\citealt{lar01}, 
see also \citealt{kun01}). 
\begin{table*}
\begin{tabular}{c|ccc|ccc}
\hline
  	  & \multicolumn{3}{c|}{Blue population}  & \multicolumn{3}{c}{Red population}\\
	  & Fraction[per cent]	& Peak colour & Width 	& Fraction[per cent]  	& Peak colour & Width \\
\hline
NGC\,3258 
	  & $76\pm3$  & $0.93\pm0.01$ & $0.10\pm0.01$	& $24\pm5$ 	& $1.13\pm0.01$ & $0.06\pm0.01$\\
NGC\,3268 
	  & $62\pm14$ & $0.98\pm0.02$ & $0.08\pm0.01$	& $38\pm7$ 	& $1.16\pm0.04$ & $0.10\pm0.02$\\
\hline
\end{tabular}
\caption{Results of two-Gaussian fittings to the colour distributions of the elliptical 
         galaxies GCSs.}
\label{tab:kmm}
\end{table*}

The total fraction of red GCs that we obtain is $24\pm5$ per cent in NGC\,3258, 
and $38\pm7$ per cent in NGC\,3268. 
\citet{rho01,rho04} found proportions of $\approx 40$ per cent of red GCs  
in their wide-field studies of NGC\,4427 and NGC\,4406, respectively.
With regard to the `intracluster' field,  
we only fit one Gaussian to the data and find $(V-I)=0.89\pm0.02, \sigma=0.09\pm0.02$ 
for the peak position and the width, respectively.

In our studies on GCSs of the Fornax cluster galaxies \citep*{dir03b,bas06b}, 
we have found that the limit between blue and red GCs is at $(C-T1) = 1.45 - 1.55$. 
By means of the comparison $(V-I)$ vs. $(C-T1)$ depicted in Fig.\,\ref{fig:checkcalib}, 
we estimate this colour limit as $(V-I) = 1.05$, that will be adopted in the rest of 
this paper and agrees with the one used by \citet{lar01}. 

\subsection{The luminosity functions}

\begin{figure}
\includegraphics[width=84mm]{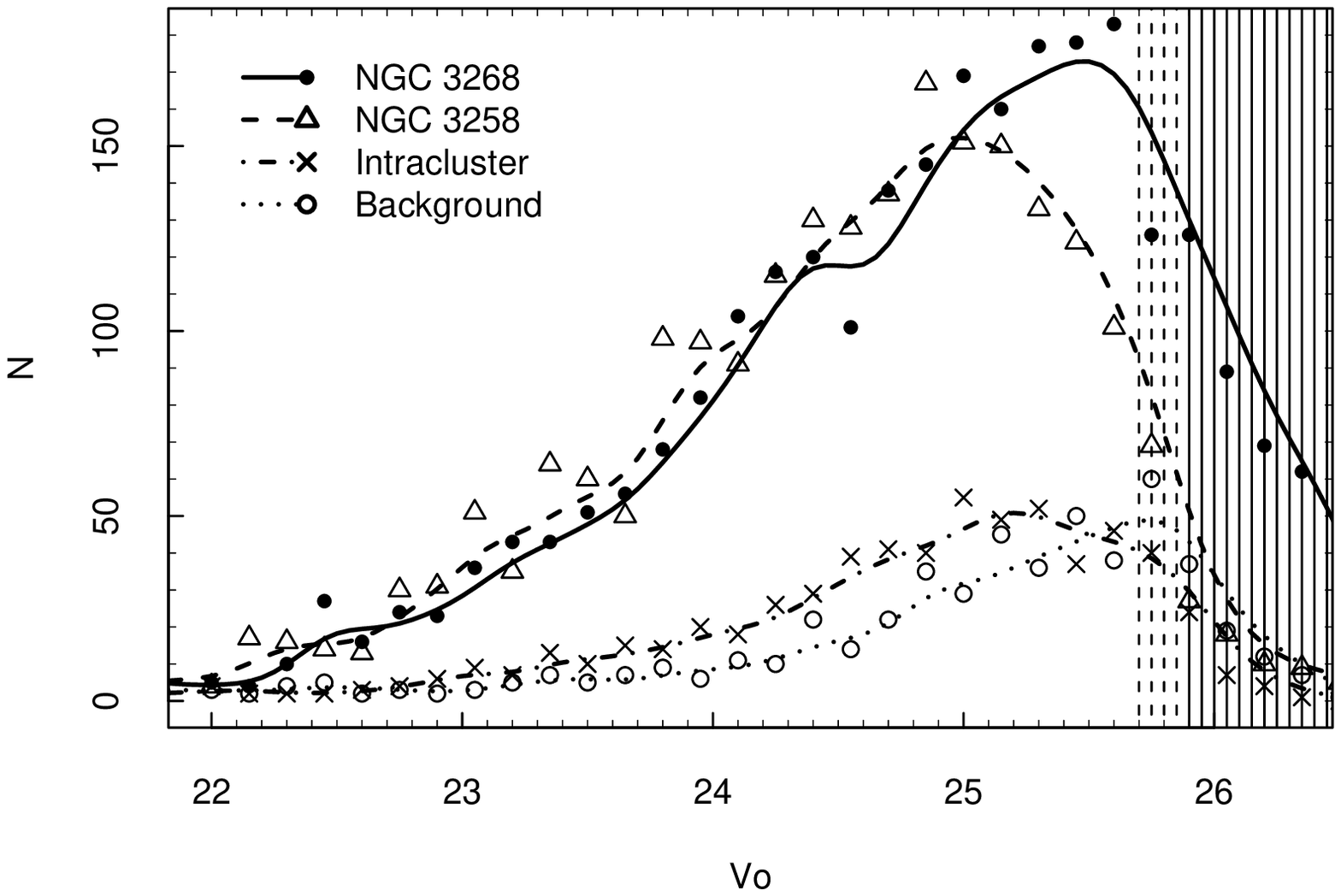}
\includegraphics[width=84mm]{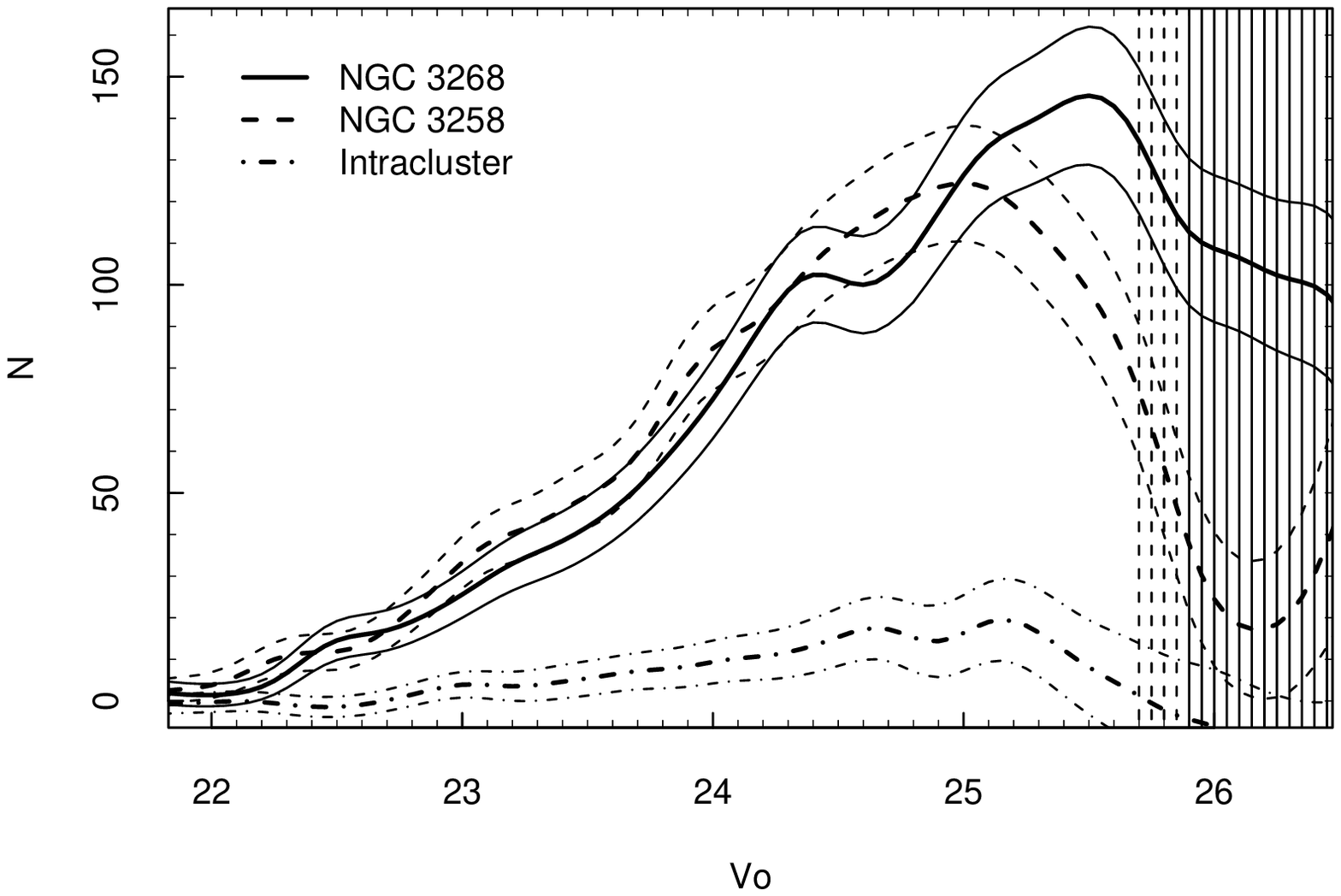}
 \caption{Upper panel: Raw LFs of all point sources in the GC colour range 
         ($0.75<(V-I<1.4$) for the four studied fields.
         The shaded areas show the luminosity ranges where the counts become uncertain 
	   because of the declining completeness
         ($V$ = 25.7 is the limit for the NGC\,3258 and
        `intracluster' fields, and $V$ = 25.9 for the NGC\,3268 field).
          Lower panel: Completeness corrected and background subtracted LFs of all 
          point sources in the GC colour range.
          The respective shaded bins are not included in the respective fitting LFs 
          (see text). Thinner lines give the corresponding errors, calculated on 
          the basis of the Poisson uncertainties of the raw and 
          background counts, and the effect of the incompleteness.}
 \label{fig:lfktraw} 
\end{figure}

The luminosity functions (LFs) of all objects in the cluster colour range 
($0.75<(V-I)<1.4$) are plotted for each of the four 
fields in the upper panel of Fig.\,\ref{fig:lfktraw} while the completeness 
corrected and background subtracted LFs are shown in the lower panel. 
In the following discussion we will only consider the GCLFs up to a 
limiting magnitude where the completeness is higher than 70 per cent, 
which is $V$=25.7 for NGC\,3258 and the `intracluster' fields. The NGC\,3268 
and the `background' fields are deeper (70 per cent completeness is reached 
at $V$=26 and $V$=25.9, respectively) but, as a background correction 
is required for the further analysis, only point sources brighter 
then $V$=25.9 in the NGC\,3268 field will be used.

For an old cluster system the LF, when counted in magnitudes, is usually close
to a Gaussian. The peak value -- turn-over magnitude (TOM) --  
corresponds to a peak in the mass distribution when counted in logarithmic
bins, which has been found to be universal for old cluster populations. 
Therefore, the GCLF can be employed for distance determinations. 
\citet{jor07} have performed the largest study of GCLFs in early-type galaxies 
to date \citep[see also][]{jor06}, within the ACS Virgo Cluster Survey. They 
have fitted two models to the LFs: a Gaussian, that is the standard model, 
and an ``evolved Schechter function'' that takes into account 
the dynamical processes that destroy the GCs, particularly the low-mass ones.  
Jord\'an et al. have shown that for bright galaxies both functions provide 
similar good fits while the largest differences arise at the low-mass 
(low-luminosity) end of the GCLF of faint galaxies. In our case, we 
are dealing with bright galaxies and do not reach the low-mass end of the 
LFs, so it seems seem appropriate to use Gaussian functions to describe the GCLFs. 
Furthermore, a t5 function has also been used to fit GCLFs (e.g. \citealt{ric03,har01})  
but, as no systematic differences in the TOMs have been reported when 
using these functions instead of Gaussians \citep{lar01}, we finally adopt the 
Gaussian model to fit the histograms, with bins of 0.15 mag. 

Several studies revealed that red and blue cluster populations have different
TOMs ( M\,87: \citealt{els96, kun99, jor02}, NGC\,4472: \citealt{puz99}, 
sample of 15 early type galaxies: \citealt{lar01}, M\,104: \citealt{spi06}) 
which is mainly due to the metallicity dependent mass-to-light ratio \citep*{ash95}. 
Hence, we study the GCLFs of the total, the red and the blue GC populations.
          
In the lower panel of Fig.\,\ref{fig:lfktraw} a TOM can be clearly seen for NGC\,3258 and 
the `intracluster' field. For NGC\,3268 the situation is slightly more complicated 
but, as it is the deepest image, a TOM can be also estimated for NGC\,3268.
The results 
are tabulated, for 
three radial subsamples, in Table\,\ref{tab:lktfits}. For the `intracluster' field no 
TOM for a red subsample has been determined due to poor number statistics.

As can be seen from Table\,\ref{tab:lktfits}, the results for the red GCs show  
larger errors, particularly in the outer radial subsample ($2\farcm3-6\arcmin$) 
where, as will be shown in Sections 5 and 6, these clusters are less numerous.  
The determined TOMs are radially independent within the errors, as has 
already been shown in other galaxies (M\,87: \citealt*{mcl94,har98,kun99,jor07}, 
the Milky Way: \citealt{har01}, M\,49: \citealt{jor07}). However, the 
TOMs depend on the colour of the GC sample: blue GCs have a brighter 
TOMs than red ones. Therefore, the red TOMs are also much poorer defined because 
of their faintness. 
The difference between the red and the blue TOMs, calculated with the results from the 
inner radial subsample where the red TOMs are better defined, is $0.53\pm0.26$  
for NGC\,3258 and $0.25\pm0.17$ for NGC\,3268. These differences agree, 
within the errors, with those reported by \citet{lar01}, but they are not 
accurate enough (particularly due to the errors in the red TOMs) to drive further 
conclusions.

\begin{table*}
\begin{tabular}{cccccccc}
\hline
	        & 		& \multicolumn{2}{c}{All GCs}  & \multicolumn{2}{c}{Blue GCs} & \multicolumn{2}{c}{Red GCs} \\
		& Radial range 	& $V$-TOM & $\sigma_V$ & $V$-TOM & $\sigma_V$ & $V$-TOM & $\sigma_V$ \\\hline
NGC\,3258 	&$0\farcm8-2\farcm3$	& $24.98\pm0.08$&$ 1.10\pm0.07$&$ 24.84\pm0.09 $&$ 1.08\pm0.09 $&$ 25.37\pm0.24 $&$ 1.17\pm0.18 $\\
		&$2\farcm3-6\arcmin$	& $24.91\pm0.12$&$ 1.09\pm0.12$&$ 24.78\pm0.12 $&$ 0.97\pm0.13 $&$ 25.59\pm0.62 $&$ 1.48\pm0.41 $\\
		&$0\farcm8-6\arcmin$	& $24.96\pm0.07$&$ 1.10\pm0.07$&$ 24.81\pm0.09 $&$ 1.04\pm0.09 $&$ 25.83\pm0.49 $&$ 1.46\pm0.28 $\\\hline
NGC\,3268	&$0\farcm8-2\farcm3$	& $25.36\pm0.09$&$ 1.21\pm0.08$&$ 25.18\pm0.13 $&$ 1.31\pm0.13 $&$ 25.43\pm0.11 $&$ 1.06\pm0.10 $\\
		&$2\farcm3-6\arcmin$	& $25.49\pm0.15$&$ 1.18\pm0.12$&$ 24.99\pm0.09 $&$ 0.96\pm0.10 $&$ 26.40\pm0.80 $&$ 1.42\pm0.37 $\\
		&$0\farcm8-6\arcmin$	& $25.35\pm0.09$&$ 1.22\pm0.08$&$ 25.06\pm0.08 $&$ 1.12\pm0.08 $&$ 25.79\pm0.18 $&$ 1.22\pm0.13 $\\\hline
Intracluster	&			& $24.79\pm0.13$&$ 0.77\pm0.14$&$ 24.87\pm0.15 $&$ 0.80\pm0.16 $&&\\\hline
\end{tabular}
\caption{V band TOMs and width of the Gaussian fittings to the GCLFs; for
	details see text.}
\label{tab:lktfits}
\end{table*}

\subsection{The distances towards NGC\,3258, NGC\,3268 and the `intracluster' field GCs}

In order to determine distances to the galaxies, we use the TOMs of the entire GC 
populations, estimated over the whole radial range, and 
adopt as universal absolute TOM $M_{V_0} = -7.46 \pm 0.18$. This universal TOM 
was determined by \citet{ric03}, as a weighted average of the TOMs of the 
Milky Way and M31 \citep*[respectively]{har01,bar01}. It is quite similar to 
the TOM recently derived for the Milky Way GCLF by \citet{jor07} 
($M_{V_0} = -7.5 \pm 0.1$).

Here are our results, where the errors in the distance moduli include the errors 
of the Gaussian fittings and the adopted universal TOM:

$(\mathrm{m}-\mathrm{M})(\mathrm{NGC\,}3258) = 32.42\pm0.19$

$(\mathrm{m}-\mathrm{M})(\mathrm{NGC\,}3268) = 32.81\pm0.20$\\ 
For NGC\,3258 and NGC\,3268 \citet{ton01} determined distance moduli of 
$32.53\pm0.27$ and $32.71\pm0.25$, respectively, which agree well with our 
measurements.  

With regard to the GC population located in the `intracluster' field, 
mainly blue GCs (see Fig.\,\ref{fig:colour1}), it seems more interesting 
to calculate its distance relative to both giant galaxies 
than to obtain an absolute estimation. 
Such relative distance may be estimated comparing the TOMs that 
are calculated using only blue globulars, over the whole radial range.
These TOMs, depicted in Table\,\ref{tab:lktfits}, 
are $V$=24.81, 25.06, and 24.87 for the NGC\,3258, NGC3268 and `intracluster' fields, 
respectively. Assuming an universal absolute TOM for the blue GC population too, 
the distance to the GCs of the `intracluster' field is in the middle of 
those of the galaxies, further than NGC\,3258 but closer than NGC\,3268.

Summarizing, we find that NGC\,3268 seems to be further away than NGC\,3258  
and that the TOM of the `intracluster' field GCs suggests that they are located 
in between both galaxies, as expected if 
most of the GCs in the `intracluster' field are part of both  
galaxies GCSs.

As a consequence of the uncertainties, particularly with regards to the 
reddening towards the Antlia cluster, in the following we will keep a 
conservative value of 30\,Mpc for the distance to the Antlia cluster (one 
arcminute will correspond to 8.7\,kpc).

%==========================================================
\section{The stellar bodies of NGC\,3258 and NGC\,3268}

\begin{figure}
\includegraphics[width=84mm]{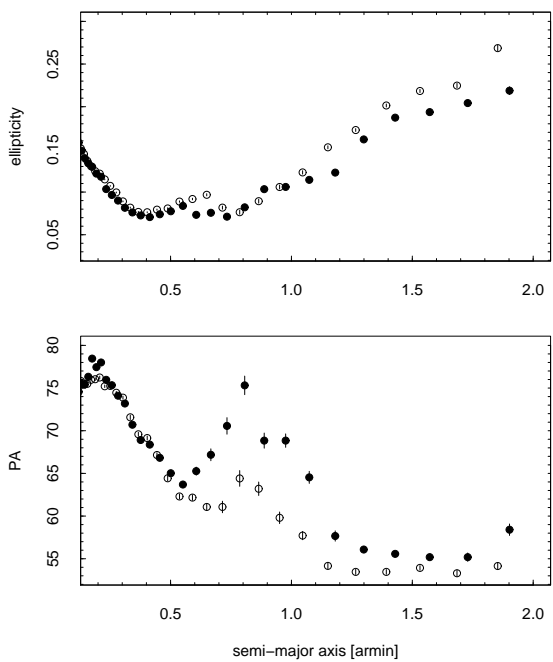}
 \caption{NGC\,3258: ellipticity (upper panel) and position angle (lower panel) 
          vs. semi-major axis, $I$-band (solid circles) and $V$-band (open circles).}
 \label{fig:azigal58}
\end{figure}

\begin{figure}
\includegraphics[width=84mm]{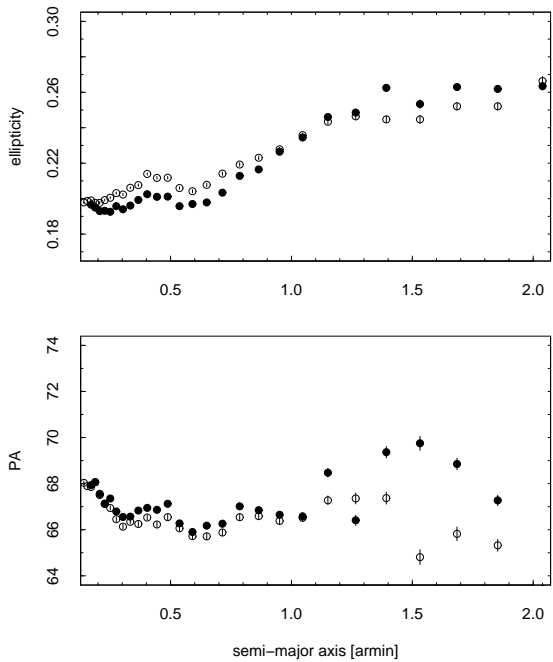}
 \caption{NGC\,3268: ellipticity and position angle plotted in the same way as 
          in Fig.\,\ref{fig:azigal58}.}
 \label{fig:azigal68}
\end{figure}

We have modeled the light of the two elliptical galaxies, as stated above, 
by means of the ELLIPSE task within IRAF. The variations of the 
ellipticity $\epsilon$ and the position angle PA against the semi-major axis, 
which result from the fits, are shown in Fig.\,\ref{fig:azigal58} and 
Fig.\,\ref{fig:azigal68} for NGC\,3258 and NGC\,3268, respectively.

From Fig.\,\ref{fig:azigal58} we can see that the ellipticity of NGC\,3258, in 
both $V$ and $I$ bands, shows a slight decline close to the center, and  
increases outwards from $\epsilon = 0.07$ at 0.5\,arcmin up to $\epsilon = 0.25$ 
at about 2\,arcmin. 
The position angle of the major-axis obtained from the isophotal analysis decreases 
from $\sim$\,75\degr close to the center down to 55\degr at about 2\,arcmin,  
with a clear peak around 0.8\,arcmin which is more evident in the $I$ band. The behaviour 
of both parameters, $\epsilon$ and PA, are in good agreement with the $BVI$ photometry 
performed by \citet*{rei94} and with our previous results from Paper\,{\sc i}.

With regard to NGC\,3268 (Fig.\,\ref{fig:azigal68}), the ellipticity increases 
steadily from $\epsilon =0.2$ at 0.1\,arcmin to $\epsilon = 0.26-0.27$ at about 2\,arcmin, 
in both $V$ and $I$ bands. 
The same results were obtained by \citet{rei94} while in the model from Paper\,{\sc i} 
the ellipticity is constant $\epsilon =0.2$ out to 2.5\,arcmin. 
The position angle remains almost constant, oscillating between 65\degr and 70\degr 
through the same radial extension, in agreement with both previous studies. 

At the centres of both galaxies, there are small dust lanes 
that can be interpreted as dusty disks. The dusty disk of NGC\,3268, with a 
diameter of 4.4\,arcsec, has already 
been mentioned in Paper\,{\sc i}. The smaller one in NGC\,3258, with a diameter of 
1.8\,arcsec, was not visible in the MOSAIC data, but has been detected by \citet{deb04}.

%==========================================================
\section{The GCS of NGC\,3258}

\subsection{Radial distribution}

\begin{figure}
\includegraphics[width=84mm]{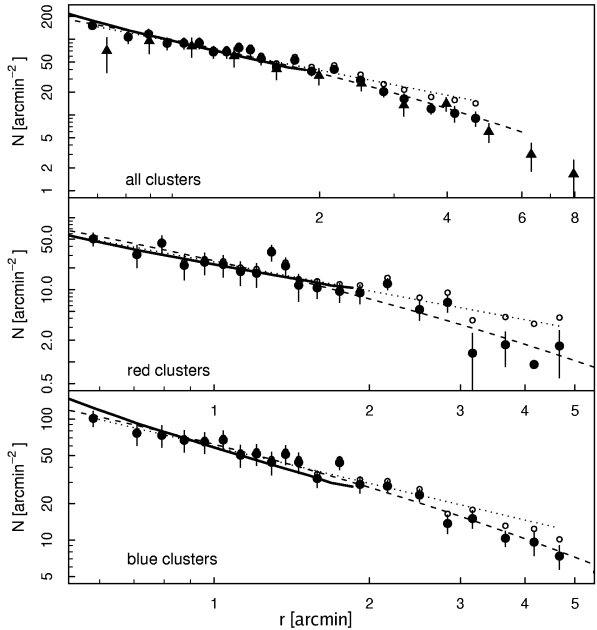}
 \caption{NGC\,3258: radial density profiles for all (upper panel), red 
	(middle panel) and blue (lower panel) GCs, brighter 
      than $V$=25. Open circles show the radial distributions uncorrected 
      for background contamination and filled circles the background-corrected 
      distributions. Triangles in the upper panel correspond to the MOSAIC data. 
      Dotted and dashed lines show the power-law and de\,Vaucouleurs 
      fits, respectively, to the background-corrected data. 
      Thick solid line represents the (arbitrarily scaled) $V$ galaxy brightness  
      profile. (Please note different horizontal and vertical scales)}
 \label{fig:radial58} 
\end{figure}

The radial density profiles of all GCs, the red ($1.05<(V-I<1.4$), 
and the blue ($0.75<(V-I<1.05$) ones, are shown in
Fig.\,\ref{fig:radial58} for the GC candidates in NGC\,3258 brighter than $V$=25.
Hence, as the TOM  for all  GCs is around $V$=25, the number density gives 
approximately half the total cluster density.  
In all cases, the errors of the background-corrected distributions include 
the Poisson uncertainties of the raw and background counts, and the effect of 
the incompleteness. The profile obtained from the MOSAIC data has been included in the 
upper panel, together with the VLT ones, as an additional check of the consistency 
between both observational sets. 
Typically GC density profiles are either fitted by power-laws 
($r^{-\alpha}$) or de\,Vaucouleurs profiles ($\exp(-a(r^{0.25}-1))$).
Both fits are plotted in Fig.\,\ref{fig:radial58}, which shows that   
de\,Vaucouleurs profiles provide better fits for all the GC selections. 
All the fits were performed within the range 0.5--5\,arcmin and the results are 
depicted in Table\,\ref{tab:radialdistr}. 

The exponents of the power-law and de\,Vaucouleurs fits show that the red clusters present a 
steeper radial profile than the blue clusters, being more concentrated towards the centre. 
A similar result, in the sense that  a de\,Vaucouleurs profile provides a better fit, 
has been found for other giants like, for instance, NGC\,4406 \citep{rho04} and 
NGC\,4472 \citep{rho01} in Virgo, or NGC\,1399 \citep{bas06a} in Fornax.  

The radial density profile for the red clusters depicted in Fig.\,\ref{fig:radial58}, 
has a rather constant and low density for r $>$ 3 arcmin (1.4 GCs\,arcmin$^{-2}$, i.e., 
about 50 per cent of the background level estimated for the red GCs color range), 
suggesting that it is close to reach   
an end, while the blue clusters clearly extend further than the NGC\,3258 field. 

We have modeled the light of the two elliptical galaxies, as stated above, 
by means of the ELLIPSE task within IRAF. The $V$ galaxy light profile included in the plots
is hardly distinguishable from the radial density profile for all and red GCs, while slight 
differences are detectable with respect to the blue cluster profile. 
 
\begin{table*}
\begin{tabular}{ccc}
\hline
 & power-law & de\,Vaucouleurs \\
\hline
 NGC\,3258 &  \\
 all clusters 	& $(83.6\pm2.2)\,r^{-1.10\pm0.06}$ & $(87.6\pm7.5)\,\mathrm{e}^{(-4.8\pm0.2)\,(r^{0.25}-1)}$\\
 red clusters 	& $(24.2\pm1.6)\,r^{-1.33\pm0.16}$ & $(25.2\pm6.3)\,\mathrm{e}^{(-6.4\pm0.5)\,(r^{0.25}-1)}$\\
 blue clusters 	& $(59.2\pm1.4)\,r^{-1.01\pm0.06}$ & $(61.9\pm5.2)\,\mathrm{e}^{(-4.3\pm0.2)\,(r^{0.25}-1)}$\\
\hline
 NGC\,3268 &  \\ 
 all clusters 	& $(70.6\pm1.9)\,r^{-1.50\pm0.09}$ & $(68.4\pm7.4)\,\mathrm{e}^{(-4.7\pm0.2)\,(r^{0.25}-1)}$\\
 red clusters 	& $(35.1\pm1.8)\,r^{-1.76\pm0.18}$ & $(33.8\pm5.2)\,\mathrm{e}^{(-5.7\pm0.3)\,(r^{0.25}-1)}$\\
 blue clusters 	& $(35.2\pm1.4)\,r^{-1.26\pm0.12}$ & $(33.8\pm4.9)\,\mathrm{e}^{(-3.9\pm0.3)\,(r^{0.25}-1)}$\\
\hline
\end{tabular}
\caption{Fits to the radial number density (objects\,arcmin$^{-2}$) of the whole GCS and two subsamples,  
         up to a limiting magnitude $V$=25, and corrected for contamination and incompleteness.
         The contribution of the background is relatively low (4.3 objects\,arcmin$^{-2}$ for the whole GC sample). 
}
\label{tab:radialdistr} 
\end{table*}

\subsection{Azimuthal distribution}

The ellipticity of the GCS can be determined studying the azimuthal density 
distribution of GC candidates with respect to the azimuthal angle, that is, a 
position angle measured from north to east. An elliptical GCS causes 
sinusoidal counts in this diagram: the ellipticity $\epsilon$ and the number density
along the major and the minor axis ($\mathrm{N}_a, \mathrm{N}_b$) are related via
$\epsilon = 1-\left(\mathrm{N}_b/\mathrm{N}_a\right)^{1/\alpha}$, where $\alpha$ is
the exponent of the radial density distribution ($r^{-\alpha}$). 

\begin{figure}
\includegraphics[width=84mm]{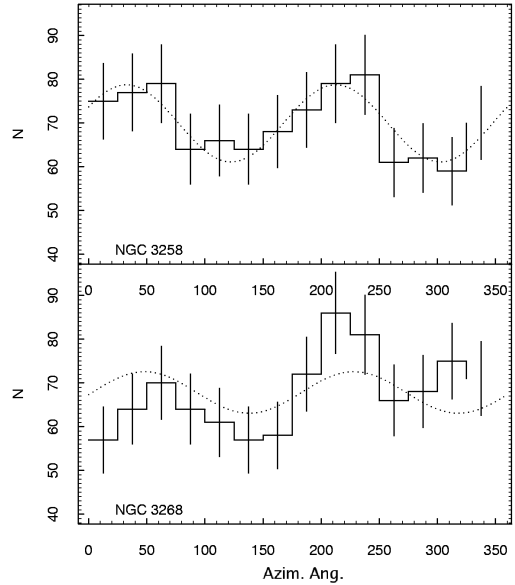}
 \caption{Azimuthal number distributions of all cluster candidates 
       brighter than $V=25.7$ within 0.5--2\,arcmin for NGC\,3258 
      (upper panel) and for NGC\,3268 (lower panel). The dotted lines 
	show the sinusoidal fits.
}
 \label{fig:azi}
\end{figure}

In the upper panel of Fig.\,\ref{fig:azi} the azimuthal distribution of the NGC\,3258 
GCS is shown for clusters within 0.5--2\,arcmin. A sinusoidal fit to the 
data results in an ellipticity $\epsilon = 0.21\pm0.03$ and a PA of the major 
axis $\mathrm{PA}=32\degr\pm5\degr$, which are in excellent agreement with those obtained 
in Paper\,{\sc i}. 

\subsection{Colour distribution}

\begin{figure}
\includegraphics[width=84mm]{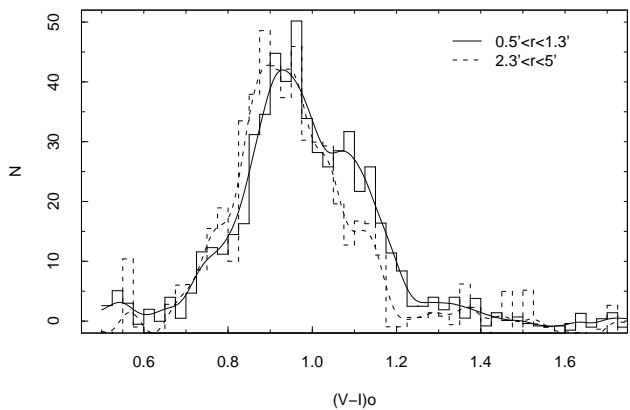}
 \caption{Colour distribution of the GCS around NGC\,3258 for an	
	inner (solid line) and an outer (dashed line) cluster sample. 
}
 \label{fig:colour58}
\end{figure}

We have shown that red and blue clusters have different radial distribution, hence 
the colour distribution shows a radial dependence which is 
shown in Fig.\,\ref{fig:colour58}. In particular, the presence of red GCs is less 
noticeable in the outer cluster sample. We have fitted two Gaussians to the 
histogram data.  
We find for the inner sample ($0.5 < r < 1.3$ \,arcmin):
\begin{eqnarray*}
&&(V-I)_\mathrm{peak}(\mathrm{blue})=0.94 \pm 0.01,~\sigma(\mathrm{blue})=0.09\pm0.01 \\
&&(V-I)_\mathrm{peak}(\mathrm{red})=1.13 \pm 0.01,~\sigma(\mathrm{red})=0.06\pm0.01
\end{eqnarray*}
and for the outer sample ($2.3 < r <5.0$ \,arcmin):
\begin{eqnarray*}
&&(V-I)_\mathrm{peak}(\mathrm{blue})=0.93 \pm 0.01,~\sigma(\mathrm{blue})=0.10\pm0.01 \\
&&(V-I)_\mathrm{peak}(\mathrm{red})=1.14 \pm 0.01,~\sigma(\mathrm{red})=0.02\pm0.01
\end{eqnarray*}
No radial dependence of the peak positions can be observed up to our limit in 
galactocentric radius, that is, about 45\,kpc. This result is consistent with the
observations in other ellipticals within similar radial ranges \citep*{lar01,rho01,dir05}. 
In this way, the radial dependence of the colour distribution may be basically explained 
by the different proportions of the blue and red subpopulations. 

We have also tested the colour distribution within different luminosity ranges: 
bright ($21.5<V<23.5$), intermediate ($23.5<V<24.5$) and 
faint ($24.5<V<25.5$) samples (the figure is not shown because it is 
similar to the NGC\,3268 one depicted in Fig.\,\ref{fig:colour68a}). 
For the intermediate and faint samples it is possible to fit two Gaussians. 
These results show that the blue and red peaks for these luminosity ranges agree 
with the colours obtained for all GCs together, within the errors (Table\,\ref{tab:kmm}). 
However, the red peak of the fainter sample is redder than the red of the intermediate 
sample ($V-I = 0.93 \pm 0.02$ / $1.09 \pm 0.04$, and $V-I = 0.92 \pm 0.02$ / $1.17 \pm 0.04$, 
for blues/reds in the intermediate and faint samples, respectively). 

For the brighter sample, it is not possible to fit two Gaussians but only one. 
The peak colour obtained ($V-I = 0.96 \pm 0.01$) is somewhat intermediate between 
both GC populations.
This behaviour has already been detected in other ellipticals that 
dominate galaxy clusters like, e.g., NGC\,1399 \citep{ost98,dir03b}, and the sample 
of eight brightest cluster galaxies from \citet{har06}.

%========================================================== 
\section{The GCS of NGC\,3268}

\subsection{Radial distribution}

\begin{figure}
\includegraphics[width=84mm]{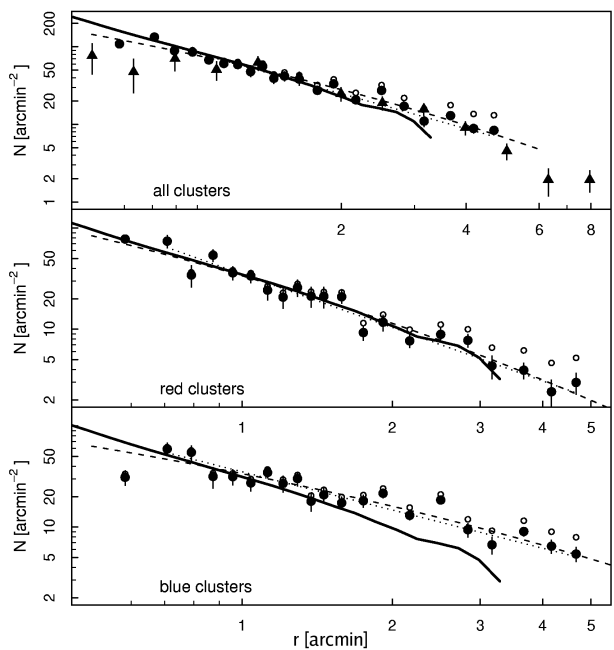}
 \caption{NGC\,3268: radial density profiles plotted in the same way as in 
          Fig.\,\ref{fig:radial58}.}
 \label{fig:radial68} 
\end{figure}

Fig.\,\ref{fig:radial68} shows the radial density profiles of all GCs, red  
and blue ones, for the candidates brighter than $V$=25 in NGC\,3268. In this case, 
the profiles can be well described by power-laws or by de\,Vaucouleurs profiles. 
All fits were performed within the 
range 0.6--5\,arcmin and the results are depicted in Table\,\ref{tab:radialdistr}.
The profile obtained from the MOSAIC data for this GCS has been included in the 
upper panel, where the difference between the VLT and MOSAIC profiles in the inner region 
is due to radial incompleteness effects within 0.7\,arcmin (see fig. 6 in Paper\,{\sc i}). 

The slopes of all de\,Vaucouleurs fits for all GCs (red and blue) agree 
within the errors with those of the NGC\,3258 fits.  
As in the NGC\,3258 GCS, red clusters have a more concentrated distribution than  
blue clusters.
It is apparent 
from Fig.\,\ref{fig:radial68} that the innermost point of the blue GCs profile    
deviates from the expected position, which is probably due to an underestimation 
of the completeness correction.

Neither the blue or red GC radial density profiles show any feature that 
can be understood as the spatial limit of the GCS. However, the red profile shows 
that the zero density level will be reached at a galactocentric radius 
slightly larger than 5\,arcmin. 

The galaxy surface luminosity profile shown in Fig.\,\ref{fig:radial68} is well 
traced by the red GCs, but the slope is clearly different from that of the blue 
GCs profile.  

\subsection{Azimuthal distribution} 

The azimuthal number counts of the GCs within 0.5--2\,arcmin are shown in the 
lower panel of Fig.\,\ref{fig:azi}, from which we derive an ellipticity 
$\epsilon = 0.09\pm0.05$ and a position angle $\mathrm{PA}=48\degr\pm20\degr$. 
The errors in both parameters are large because the fit is affected by 
a clear excess in the GC azimuthal distribution, at azimuthal angles between 
200--250\degr. This excess appears in coincidence with one of the maxima of the 
sinusoidal fit, and at azimuthal angles that correspond to the direction towards 
NGC\,3258. 
 
It is interesting to note that on the sky, the position angle 
with origin in NGC\,3258 that points to the direction of NGC\,3268, 
is 39\degr. So we confirm the results from Paper\,{\sc i}, that 
both GCSs are elongated in a direction close to an axis joining 
the two galaxies. 

\subsection{Colour distribution}  

\begin{figure}
\includegraphics[width=84mm]{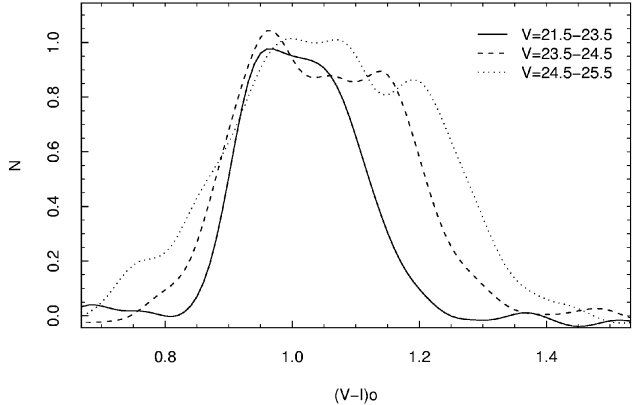}
 \caption{Colour distribution of the GCS of NGC\,3268 for three	
	different luminosity ranges. The distributions are arbitrarily scaled. 
} 
  \label{fig:colour68a}
\end{figure}

The GC colour distribution is shown in Fig.\,\ref{fig:colour68a} 
for three luminosity intervals, within the radial range 0.7--2.3\,arcmin. 
As for NGC\,3258, 
the brightest GCs show a unimodal colour distribution 
while fainter GCs have bimodal distributions. 
The distribution also extends to redder colours as we consider fainter GCs. 
 
The two-Gaussian fits show that the blue and red peaks, for the intermediate 
and faint luminosity ranges, agree with the colours obtained for all GCs together, 
within the errors (Table\,\ref{tab:kmm}). However, the red peak of the fainter sample 
is clearly redder than that of the intermediate sample ($V-I = 0.96 \pm 0.01$ / 
$1.14 \pm 0.02$, and $V-I = 1.02 \pm 0.02$ / $1.23 \pm 0.02$, for blues/reds in the 
intermediate and faint samples, respectively). Similar trends are present in 
the NGC\,3258 GCS. The peak colour obtained from the single-Gaussian fit to the 
brightest sample ($V-I = 1.01 \pm 0.01$) is roughly in between both subpopulations. 

The colour distributions within different galactocentric radii have also 
been tested within 
an inner and outer samples as for NGC\,3258. We confirm the radial 
dependence of the colour distribution 
because the fraction of red GCs present in the outer sample is smaller than 
in the inner one. 
The colours of the blue/red peaks roughly agree, within the errors, with the 
ones estimated for the whole GC population, but in this case the peaks of 
the inner sample are redder than those of the outer group ($V-I = 1.02 \pm 0.01$ / 
$1.21 \pm 0.02$, and $V-I = 0.96 \pm 0.02$ / $1.14 \pm 0.04$, for blue/red 
peaks in the inner and outer samples, respectively).  

%==========================================================
\section{Discussion} 

\subsection{The brightest GC candidates}

The principal property of the colour distribution of GCs, the bimodality, has been previously 
found to be absent among the bright GCs of NGC\,1399 \citep{ost98,dir03b} and M87 \citep{str06}. 
\citet{har06} found the same in their ACS photometry of several central giant ellipticals, 
including our Antlia galaxies, while this point remained unclear in our previous Washington 
photometry. Unimodal colour distributions, 
seem to apply to clusters brighter than approximately $M_V = -10$. 
Our present photometry reiterates on this finding, shifting the limit between bimodality and 
unimodality even a bit lower to $M_V = -9$.

It is plausible (and discussed in the literature, see the reviews of \citealt{ric06} and 
\citealt{bro06}) that there are several possible formation channels for creating a population of 
very bright clusters. Stripped galactic nuclei \citep{bas94}, former Blue Compact Galaxies, 
but also ``normal'' formation of massive clusters, perhaps through the merging of 
smaller subclusters \citep{fel05} are viable candidates or formation histories, 
which in the center of a galaxy cluster might occur more frequently than in less dense 
environments.

\subsection{Total GC populations and specific frequencies}

We can estimate the GC populations performing a numerical integration of the de\,Vaucouleurs  
radial density profile, which includes GCs brighter than $V$=25, and doubling the result 
according to the TOMs of the GCLFs. 
The external limiting radius of the GCSs is taken as $r$ = 10\,arcmin following Paper\,{\sc i}. 
In this way, the total GC population of NGC\,3258 amounts to $N_{GC} = 6000 \pm 150 $ 
and that of NGC\,3268 $N_{GC} = 4750 \pm 150$. In NGC\,3258, the red GCs are clearly less 
numerous than the blues, being the ratio of blues to reds $N_b / N_r = 4.1 $, which is close 
to the ratio estimated from the colour distribution $N_b / N_r = 3.2 \pm 0.7$, within 
the errors (see Table\,\ref{tab:kmm}). 
The ratio of blue to red GCs around NGC\,3268 is $N_b / N_r = 1.7$, also close 
to the ratio derived from the colour distribution $N_b / N_r = 1.6 \pm0.8$ 
(Table\,\ref{tab:kmm}). 

In order to calculate the specific frequency $S_{N}$ \citep[as defined by][]{har81} 
of both GCSs, we need the $V$-band absolute magnitudes of the host galaxies and number 
of GCs estimated over the same galactocentric distance. The $R$-band integrated luminosities  
of the galaxies up to $r$ = 4\,arcmin can be obtained from Paper\,{\sc i} and corrected 
by absorption in $R$ applying the relation from \citet{rie85} $A_R/A_V = 0.75$ 
and the $E(B-V)$ colour excesses depicted in Table\,\ref{tab:fields}. We cannot estimate 
global $S_{N}$ due to the large uncertainties of the integrated luminosity at larger 
radii, but we are able to improve the $S_{N}$ from Paper\,{\sc i} (also within $r$ = 4\,arcmin) 
through a more precise determination of the number of GCs. By means of the individual 
distance moduli estimated in this paper and the colour index $(V- R) = 0.7$ (Paper\,{\sc i}) 
for both galaxies, we obtain absolute luminosities within 4\,arcmin $M_{V}= -21.5 \pm 0.3 $ 
for NGC\,3258, and $M_{V}= -22.2 \pm 0.3 $ for NGC\,3268, where the errors 
are calculated with the errors of the apparent magnitudes and of the distance moduli. 
The corresponding GC populations, up to the same radius, are estimated by numerical 
integration of the radial density profiles. Finally, the specific frequencies 
within $r$ = 4\,arcmin are $S_{N} = 8.7 \pm 2.2 $ for NGC\,3258, and  $S_{N} = 3.7 \pm 0.9$ 
for NGC\,3268. Due to the limited radial range of these calculations, these $S_{N}$ should 
be taken as indicative values.

\subsection{Distances and specific frequencies}
 
The derived distances indicate that NGC\,3258 is located
somewhat in the foreground with respect to NGC\,3268. The TOMs, taken at face value, 
suggest a difference
of 6 Mpc, while the SBF distances are different by 3 Mpc, although the uncertainties
would not exclude the same distance.  
An additional argument is that the `intracluster' field independently
reveals a TOM intermediate between NGC\,3258 and NGC\,3268 which we expect if the outskirts of the
respective GCSs are projected onto each other. 

This interpretation is also not without oddities. NGC\,3258 has an unusually high specific frequency
considering its brightness and, if really in the foreground, would be in a relatively poor environment, at least
poorer than that of NGC\,3268. Such a high frequency is, however, not unique as we see from the case 
of NGC\,4636 \citep{dir05}, but extraordinary. A distance modulus of 0.4 mag 
higher would decrease its specific frequency by a factor of 1.4, making it more common. On the other hand,
its TOM is very well sampled, better than that of NGC\,3268, so we do not think that the TOM is grossly
erroneous. 

A few remarks to the group around NGC\,3268: The radial velocities of the three neighbour galaxies 
that are listed in Paper\,{\sc i} (NGC\,3269, NGC\,3271, NGC\,3267; these are the  only ones for 
which radial velocities are available), are consistently 
higher by about 1000 km/s. The morphological appearance suggests NGC\,3268 to be the central 
galaxy of that subgroup of the Antlia cluster. These radial velocities, however, let it appear 
improbable that NGC\,3268 is at rest with respect to that group. Its ``normal'' specific frequency also      
would not qualify it as a ``central'' galaxy. 

More insight into the structure of the Antlia cluster can only be expected by a radial velocity survey.

\subsection{Intracluster globular clusters?}

With regard to the presence of GCs inside galaxy clusters
that are not bound to individual galaxies, 
observational evidence of their existence have been presented, among others, 
by \citet{min98}, \citet{kis99} \citet{bas03}, \citet{jor03}, and \citet{wil07}.  
Besides, numerical simulations on their formation and properties have 
been performed by \citet{yah05}.

\begin{figure}
\includegraphics[width=84mm]{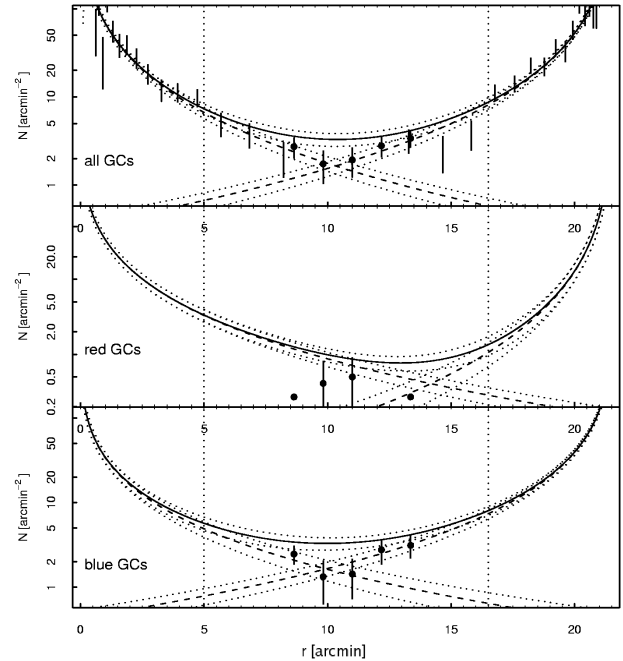}
 \caption{Radial densities along the axis connecting both Antlia galaxies, 
        where r is the galactocentric radius measured from NGC\,3268 centre.  
        The fits to the radial density profiles of all, red, and blue GCs 
	around NGC\,3258 (right side) 
        and NGC\,3268 (left side), are extrapolated and compared to the GC density 
        in the `intracluster' field. 
        Individual GCS profiles are shown with dashed lines and the combined one with 
        a solid line. Dotted lines close to the profiles give their respective 
        errors, and vertical dotted lines show the limits of the FORS1 NGC\,3258 and 
        NGC\,3268 fields.
        The `intracluster' field is divided into 5 radial ranges whose densities are 
        shown with filled circles. The MOSAIC measurements 
        are included in the upper panel, as vertical solid lines corresponding to 
        the errorbars. Please note that vertical scales are different. 
         } 
 \label{fig:intergal} 
\end{figure}

In the previous sections we have derived the cluster radial density distribution and
fitted de\,Vaucouleurs laws to the GCSs of NGC\,3258 and NGC\,3268.
We now take these fits, extrapolate them beyond the observed 5\,arcmin, 
and compute the combined cluster density along 
the connecting axis between the two galaxies. The results for all GCs, as well 
as for reds and blues, are shown in Fig.\,\ref{fig:intergal}, where NGC\,3258 
and NGC\,3268 would be on the right and left sides, respectively.
  
The `intracluster' field is located on this connecting axis, and we aim at comparing 
the observed densities in this field with the combined extrapolated profile, along this 
connecting line. 
For this reason, the GC densities for five different radial ranges within the 
`intracluster' field, completeness corrected and background subtracted, are determined 
taking a limiting magnitude $V$ = 25. The observed densities follow closely the individual 
extrapolated density profiles (filled circles in Fig.\,\ref{fig:intergal}). However, 
number statistics are low as all densities in this field, observed and extrapolated, 
are below 5\,GCs\,arcmin$^{-2}$. 
We scaled the MOSAIC densities to those of the VLT fields, which are shown in the 
upper panel of Fig.\,\ref{fig:intergal}. 
On the `intracluster' field, the combination of the extrapolated fitted functions 
seems to predict a higher cluster density than the observed one. 

If there were some intracluster GCs 
one would expect the observed density in the `intracluster' field to be 
larger than just the sum of the individual GCS extrapolated profiles. The contribution 
of red GCs is almost negligible, so the analysis is performed basically on the blue 
ones. Along the axis connecting both galaxies, the range of the observed (blue GC) 
densities is $1.3\pm 0.5 - 3.1\pm 0.6$\,arcmin$^{-2}$ while the range of combined (blue GCs) 
extrapolated densities, within the same radii, is $3.3 \pm 0.6 - 4.2 \pm 0.6$\,arcmin$^{-2}$. 
Though we are dealing with poor number statistics these results suggest that not only the 
observed densities are not larger than the predicted ones, but they even tend to be 
smaller. 
 
So far, we do not find from our data any conclusive evidence of the existence 
of intracluster GCs in the region between NGC\,3258 and NGC\,3268. However, it 
should be taken into account that the `intracluster' field, where there are  
GCs contributed by both GCSs, 
has not turn into a proper place to seek for them.    
We are undertaking a kinematic study of the GC and galaxy content of the Antlia 
cluster, that will probably help to detect their presence in this cluster.  

%==========================================================
\section{Summary and Conclusions}

On the basis of FORS1/VLT ($V,I$) images we have performed an analysis of 
the globular cluster systems of NGC\,3258 and NGC\,3268, 
the dominant elliptical galaxies of the Antlia galaxy cluster.
Our first study of these GCSs was based on Washington ($C,T1$) photometry 
and wide-field MOSAIC images (Paper\,{\sc i}), which did not reach the turn-over 
magnitude (TOM) of the globular cluster luminosity function (GCLF). 
Here we summarize the results and conclusions:

\begin{itemize} 
\item[-]The TOMs of the red, the blue, and the total populations are obtained by fitting
Gaussians to the respective GCLFs.
The distance moduli, obtained from the entire GC sample, are $(\mathrm{m}-\mathrm{M})
= 32.42\pm0.19$ for NGC\,3258 and $(\mathrm{m}-\mathrm{M}) = 32.81\pm0.20$ for NGC\,3268,
which are in good agreement with those obtained by \citet{ton01} via the SBF method. 
The TOMs of the blue GCs are on the average $<\Delta{V}>$ = 0.4 mag brighter 
than those of the red ones.  

\item[-] The GCLF was independently determined for a field between NGC\,3258 and NGC\,3268.
We could measure a TOM of $V= 24.87\pm0.15$ (blue GCs), intermediate between the TOMs of the 
bright galaxies. This supports the view that we are observing
the overlapping of the two GCSs. The actual number density is even somewhat lower than from
one would expect by the extrapolation of the number density profiles determined near the
host galaxies. We therefore found no evidence of the presence of intracluster GCs in the
field between the ellipticals.

\item[-]
The total GC populations
are about $6000\pm150$\,GCs in NGC\,3258 and $4750\pm150$\,GCs in NGC\,3268  while
the extent of both GCSs is at least 10\,arcmin (about 90 kpc). If the relative distances
are correct, this corresponds to specific frequencies of $S_N = 8.7 \pm 2.2$ for NGC\,3258
and $S_N = 3.7 \pm 0.9$ for NGC\,3268.

\item[-]
Other findings from Paper\,{\sc i} like the bimodal colour distribution or the
azimuthal distributions have been confirmed. The galaxy light profiles match more 
closely the red GCs radial density profiles.  A point not addressed in Paper\,{\sc i} 
is the unimodal colour distribution of the brightest clusters.

\end{itemize}

The strongest indication so far for the spatial proximity of NGC\,3268 and NGC\,3258 was the
common radial velocity of 2800 km/s. But the GCLF distance moduli rather suggest that NGC\,3268 
is located somewhat in the background, separated from NGC\,3258 by several Mpc.
The unusually high specific frequency of NGC\,3258 is, however, a rare finding and 
would ease with a distance equal to that of NGC\,3268. A further confirmation would
therefore be required. 
Much evidence suggests that Antlia is  a cluster in a very early stage of 
dynamical evolution, but the spatial and dynamical relationship of its components
are still unclear. We started a kinematic study intending to  bring more clarity in
the understanding of the apparently complex structure of Antlia.

%==========================================================
\section*{Acknowledgments}
BD and TR gratefully acknowledge support from the Chilean Center for Astrophysics 
FONDAP No. 15010003. LPB is grateful to the Astronomy Group at the Concepci\'on University, 
for financial support and warm hospitality during part of this research. This 
work was also funded with grants from Consejo Nacional de Investigaciones Cient\'{\i}ficas 
y T\'ecnicas de la Rep\'ublica Argentina, Agencia Nacional de Promoci\'on Cient\'{\i}fica 
Tecnol\'ogica and Universidad Nacional de La Plata (Argentina).

\label{lastpage}

\end{document}